\documentclass{emulateapj}

\usepackage{graphicx}
\usepackage{epstopdf}

\newcommand{\kepler}{{\it Kepler\/ }}
\newcommand{\keplers}{{\it Kepler's\/ }}

\newcommand\moire{moir\'{e}\ }

\slugcomment{accepted for \apjl\ Letters, 2009 December 21}

\shorttitle{INSTRUMENT PERFORMANCE IN {\it KEPLER'S} FIRST MONTHS}
\shortauthors{Caldwell et al.}

\begin{document}

\title{Instrument Performance in {\em Kepler}'s First Months}


\author{Douglas A. Caldwell\altaffilmark{1}}
\affil{SETI Institute/NASA Ames Research Center, MS 244-30, Moffett Field, 
CA 94035}

\author{ Jeffery J. Kolodziejczak}
\affil{NASA Marshall Space Flight Center, VP60, Huntsville, AL 35812 }

\author{Jeffrey E. Van Cleve, Jon M. Jenkins, Paul R. Gazis}
\affil{SETI Institute/NASA Ames Research Center, MS 244-30, Moffett Field, 
CA 94035}

\author{Vic S. Argabright, Eric E. Bachtell}
\affil{Ball Aerospace \& Technologies Corp.,1600 Commerce Street, Boulder, 
CO 80301}

\author{Edward W. Dunham}
\affil{Lowell Observatory, 1400 West Mars Hill Road, Flagstaff, AZ 86001}

\author{John C. Geary}
\affil{Smithsonian Astrophysical Observatory, 60 Garden St., Cambridge, MA 
02138}

\author{Ronald L. Gilliland}
\affil{Space Telescope Science Institute, 3700 San Martin Drive, Baltimore, MD 
21218}

\author{Hema Chandrasekaran, Jie Li, Peter Tenenbaum, Hayley Wu}
\affil{SETI Institute/NASA Ames Research Center, MS 244-30, Moffett Field, 
CA 94035}

\author{William J. Borucki, Stephen T. Bryson, Jessie L. Dotson, Michael R. Haas, 
David G. Koch}
\affil{NASA Ames Research Center, MS 244-30, Moffett Field, CA 94035}

\altaffiltext{1}{\tt{douglas.daldwell@nasa.gov}}

\begin{abstract}
The {\it Kepler Mission} relies on precise differential photometry to detect the 80 
parts per million (ppm) signal from an Earth--Sun equivalent transit. 
Such precision requires superb instrument stability on 
time scales up to  $\sim$2 days and systematic error removal to better than 20 
ppm. To this end, the spacecraft and photometer underwent 67 days of 
commissioning, which included several data sets 
taken to characterize the photometer performance. Because \kepler has no 
shutter, we took a series of dark images prior to the dust cover ejection, 
from which we measured the bias levels, dark 
current, and read noise. These basic detector properties are essentially 
unchanged from ground-based tests, indicating that the photometer is 
working as expected. Several image artifacts have proven more 
complex than when observed during ground testing, as a result of 
their interactions with starlight and the greater 
thermal stability in flight, which causes 
the temperature-dependent artifact variations to be on the timescales of 
transits. Because of \keplers 
unprecedented sensitivity and stability, we have also seen several unexpected 
systematics that affect photometric precision. We are using 
the first 43 days of science data to characterize these effects and to develop 
detection and mitigation methods that will 
be implemented in the calibration pipeline. Based on early testing, we expect 
to attain \keplers planned photometric precision over 80\%--90\% of the field
of view.

\end{abstract}


\section{Introduction}

\kepler was launched on 2009 March 6, beginning a 3~1/2 
year mission to detect transiting exoplanets and determine the 
frequency of Earth-size planets in the habitable zones of solar-like stars. The 
objectives and early results of the {\em Kepler Mission} are reviewed by 
\nocite{boruckiScienceRev2010} Borucki, et al.\ (2010) and the mission design 
and overall performance are reviewed by \nocite{kochMissionDesign2010} 
Koch et al.\ (2010). Before beginning science operations, \kepler underwent a 
commissioning period to ensure it was operating correctly after the rigors of 
launch, to verify that ground-based characterizations were still valid, and to 
perform characterizations that could only be done in space
\citep{haasSciops2010}. In 
this Letter we describe the instrument characteristics relevant for 
understanding \keplers raw pixel data. In addition to 
standard CCD detector properties (\S~\ref{sec:stdccd}), we discuss the 
characterizations and 
data products resulting from \keplers unique design and operation 
(\S~\ref{sec:unique}). In \S~\ref{sec:instartifacts}, we 
describe several image artifacts that are present in \kepler data and 
discuss their impact on photometric precision. Detailed descriptions of the
photometer beyond the scope of this Letter can be found in 
\citet{argabrightSpie2008} and in the 
``Kepler Instrument Handbook'' \citep{vancleveKIH2009}.

\section{Observation Modes}
\kepler science data are available at either short cadence ($\sim$1 minute) 
for 512 targets, or long cadence ($\sim$30 minutes) for 170,000 targets. 
All science data are collected 
with an integration time of 6.02 s. 
In science collection mode, the full single integration CCD frames 
are coadded together, then
at the end of the short and long cadence period pre-specified 
pixels for each target are selected from the coadd, processed, 
and stored on board. 
Due to data storage and transmission limitations, only about 6\% of the 96 
million pixels are stored for eventual transmission to the ground. 

\begin{figure}
\hspace{-0.0in}
\includegraphics[scale=0.5,clip=true]{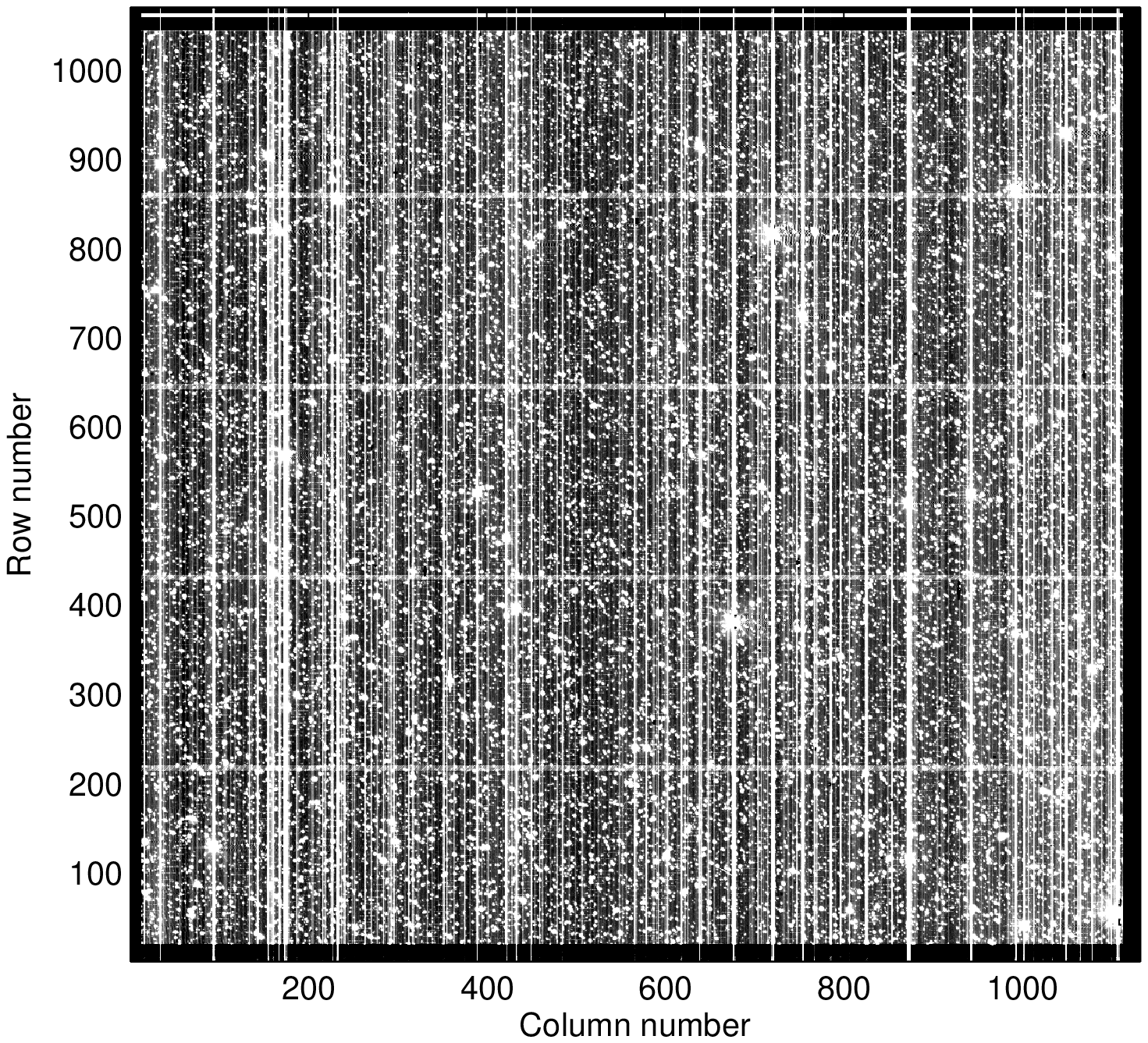}
\hspace{-0.6in}
\includegraphics[scale=0.5,clip=true]{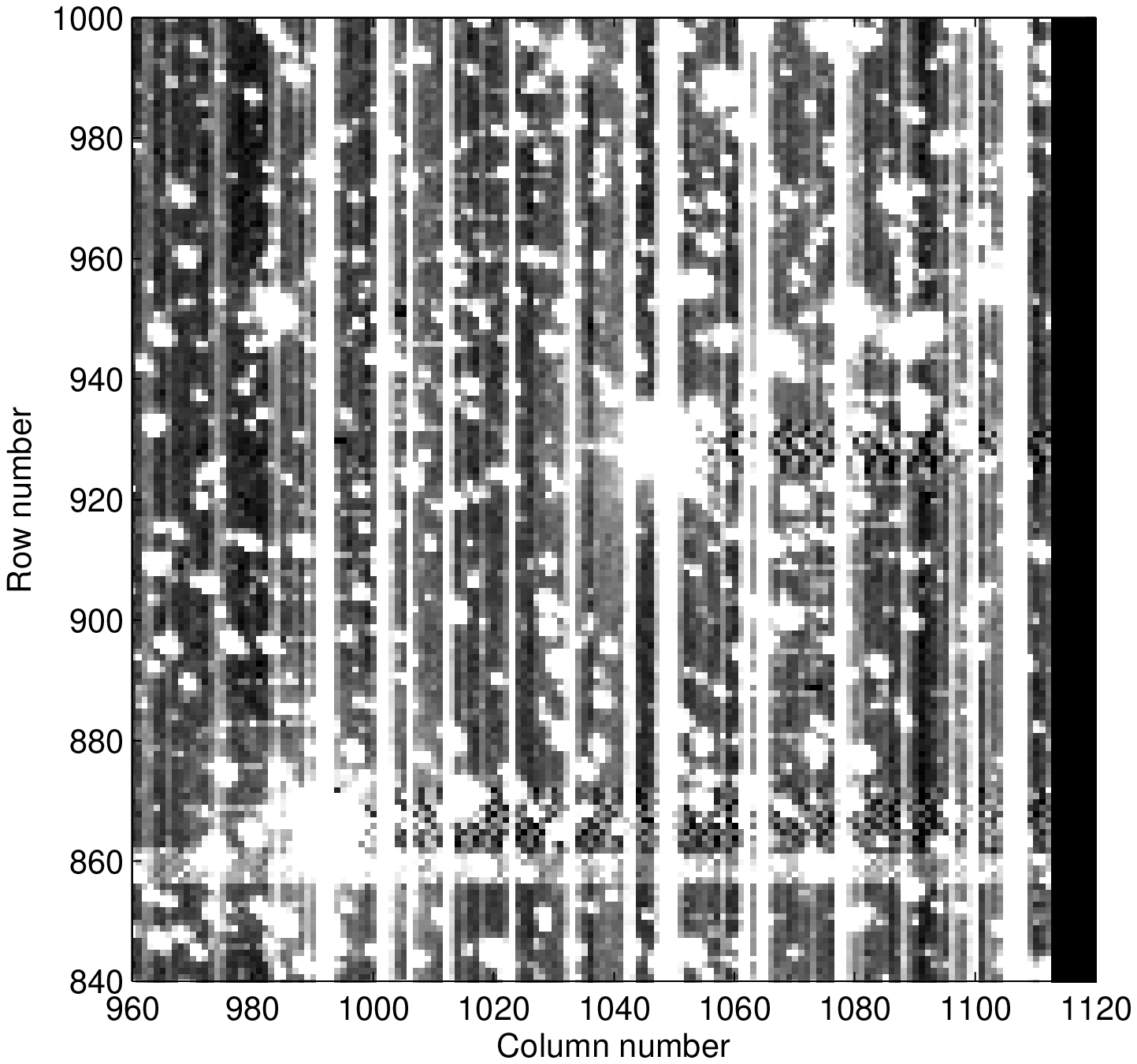}

\caption{Raw FFI of channel 17.2 (left) and a 
zoomed portion of the image (right). The left panel shows the collateral 
data regions: leading black columns (left edge), trailing black columns 
(right edge), masked smear rows (bottom), and virtual 
smear rows (top). The four bright charge injection rows can be seen 
in the virtual smear. The smear signal is visible as bright columns. 
The FGS frame-transfer clocking cross talk signals are the five 
faint horizontal bands near rows 1, 220, 430, 640, and 860. The right panel 
shows a close-up near two bright stars. The smear signal, frame-transfer 
cross talk at row 860, and trailing black beginning at Column 1112 are 
visible. The FGS parallel-transfer cross talk signals are the $\sim$16 
pixels wide segments spaced seven rows apart beginning near the lower left 
and offset toward the upper right. Scene-dependent \moire 
pattern is visible in the columns following the two bright stars in rows 865 
and 930. This commissioning FFI (KPLR2009108050033) used 270 coadds 
of 2.59 s integrations each, for a total 
exposure time of 700 s. The display uses a linear stretch from 
the $7^{\rm th}$ to $80^{\rm th}$ percentile. \label{fig:ffi} }
\end{figure}

The focal plane consists of 84 separate science readout channels 
(identified as module\#.output\#) and four fine guidance 
sensor (FGS) channels all of which are 
read out synchronously. 
Each channel has several regions available to collect calibration, or 
``collateral'' data (Figure~\ref{fig:ffi}). 
There are two sets of columns of virtual pixels: (1) 12 
columns of bias-only pixels resulting from 12 
leading pixels in the serial register (``leading black''), and 
(2) a 20 column serial over-scan region (``trailing black''). 
There are also two sets of rows of 
collateral pixels: (1) the first 20 rows, which are covered by an aluminum 
mask (``masked smear''), and 
(2) a 26 row parallel over-scan region (``virtual smear'').  
During science data collection, a coadded sum of specified columns 
of the trailing black and rows of both the masked and virtual smear are stored 
at each cadence for each channel.

Since \kepler has no shutter, we cannot take standard dark frames. Instead, 
the CCDs can be reverse-clocked so that none of the signal from the stars and 
sky reaches the CCD output, allowing us to measure the bias level 
throughout the image. 

Finally, a full frame image (FFI) mode is available, in 
which all the pixels in the focal plane are stored. FFIs are invaluable for 
examining detector properties, verifying pointing, and verifying the target 
aperture definitions; however, at 380 megabytes each, only 
a limited number can be processed and stored. 
Reverse-clocked data and FFIs are taken periodically throughout the 
mission \citep{haasSciops2010}.

\section{Detector Properties}
Each of \keplers detector channels has distinct properties 
important for understanding and analyzing the pixel data,
roughly divided into standard 
CCD detector characteristics and those unique to \keplers design. 
Where possible instrument characterization was performed on the
ground \citep{argabrightSpie2008}, though some
had to be done during commissioning.
A significant portion of commissioning was spent with the 
dust-cover on, in order to measure 
characteristics that would later be masked by star light. 
The latter half of commissioning included measurements 
of the photometer's 
point spread function, pixel response, and focal plane geometry 
parameters, as discussed in \citet{brysonPrf2010}. Commissioning 
observations resulted in a series of focal plane models that are used 
throughout the pipeline processing and are available at the data archive
\footnote{\tt{http://archive.stsci.edu/kepler/}}.

\subsection{Standard Detector Properties \label{sec:stdccd}} 

Because of the large number of photons collected from our 
typical targets, low read noise is 
not critical and the focal plane median value of 95 e$^-$ read$^{-1}$, 
or approximately 1 digital number (DN) per read, is sufficient. 
The focal plane is maintained at $-85^{\circ}$C, reducing 
dark current effectively to zero. 
The CCDs are operated to guarantee that the full-well of 1.1 
million electrons is not clipped by the 14-bit analog-to-digital converter (ADC)
range, resulting in a median gain of 112 electrons DN$^{-1}$.  
The high quantum 
efficiency (QE) back-illuminated CCDs combine with the broad bandpass so 
that stars with \kepler magnitude $\lesssim 11.3$ saturate depending on 
the field of view (FOV) location. The observed 
nonlinearity over the full range of input up to and
beyond saturation is on the order 
of $\pm 3\%$ after accounting for charge bleeding. 

Since \keplers goal is not absolute photometry, an accurate global flat field 
image is not required, but we do use a local flat, or pixel response 
non-uniformity (PRNU) map for calibration. The PRNU image 
maps each pixel's relative 
brightness variation from the local mean, expressed in percent
\citep{vancleveKIH2009}. 
The median standard 
deviation of the pixel values in the PRNU image 
across the focal plane is $0.96 \%$. A ``bad pixel'' map, 
constructed by thresholding the PRNU map to find $> 5\sigma$ outliers,
shows that $\leq 0.5\%$ of the FOV is affected 
by pixel or column defects.

Table~\ref{tab:dpp} summarizes the standard CCD detector properties 
as measured during ground testing and updated during photometer 
commissioning. 

\subsection{{\it Kepler}-Specific Instrument Properties}
\label{sec:unique}

\keplers design and operation result in several non-standard 
properties that influence the content of the raw pixels:
readout smear, charge injection, and digital data requantization.

Because there is no 
shutter, stars shine on the CCDs during readout, resulting in 
trails along columns that contain stars. 
Each pixel in a given column of the image ---including the masked and 
virtual smear rows--- receives the same smear signal 
(Figure~\ref{fig:ffi}). The column-by-column smear level in each
image is measured in the smear regions. Smear signals are typically 
small, since each pixel only ``sees" a star for the readout time (0.52 s)
divided by the total number of rows (1070); therefore smear is not
a significant contributor to photometric noise.

The science CCDs are operated with an electrical charge injection 
feature that injects charge at the top of each CCD for four consecutive 
rows at a signal level 
approximately 40\% of full well. The 
signal appears entirely in the virtual smear.
Charge injection serves the dual purpose of filling radiation 
induced traps in the CCDs and providing a stable
signal for monitoring the readout electronics, or 
``local detector electronics'' (LDE), undershoot artifact 
(\S~\ref{sec:undershoot}). 

In order to store and downlink pixel data for 170,000 targets, the CCD
output must be compressed from 23 bits pixel$^{-1}$ to 
4--5 bits pixel$^{-1}$. Because of the Poisson noise intrinsic in the data,
we can afford to requantize (after the initial analog-to-digital conversion) 
so that the effective noise due to quantization
is a constant percentage of the intrinsic noise for all signal levels. 
For higher signals with more shot noise, more ADC output 
values are mapped on to a single requantized value. 
With $\Delta_Q$ the step size for a signal with intrinsic variance
$\sigma^2_{\rm measured}$, the total variance, $\sigma^2_{\rm total}$, is
the quadrature sum of the observational noise and the quantization noise:
\begin{equation}
\label{eq:requant1}
\sigma^2_{\rm total} = \sigma^2_{\rm measured} + \Delta_Q^2/12.
\end{equation}
(Note that the variance of a uniform random variable of unit width is 1/12.)
\kepler data are requantized such that the quantization noise 
is at most 1/4 of the intrinsic noise, 
$\Delta_Q/\sqrt{12} \leq \sigma_{\rm measured}/4$, resulting in a total noise
increase of $3\%$. 
Requantization reduces the
number of bits pixel$^{-1}$ from 23 to 16 and greatly improves 
compressibility for subsequent lossless steps in the compression 
process \citep{haasSciops2010}.

\section{Instrument Artifacts}  
\label{sec:instartifacts}

Ground testing uncovered several
instrumental artifacts, each of which was investigated 
to understand the cause, impact, and cost to fix or mitigate. 
Based on reviews by the \kepler
Team and outside experts, the project dispositioned 
each of these artifacts. There were several
for which the \kepler Team and review boards decided that the potential impact to 
the mission did not warrant the risk of fixing them. These artifacts were extensively 
characterized on the ground and then again during commissioning. For 
those with the largest impact, the data processing pipeline either already
corrects for them, or corrections are under development 
\citep{jenkinsPipeline2010}.  
Table~\ref{tab:artifact} summarizes the
FOV potentially affected by each artifact. 
Values for artifact levels given below
are based on 43 days of science data collection. 

\subsection{LDE Undershoot}
\label{sec:undershoot}
\begin{figure}
\hspace{-0.0in}
\includegraphics[scale=0.5,clip=true]{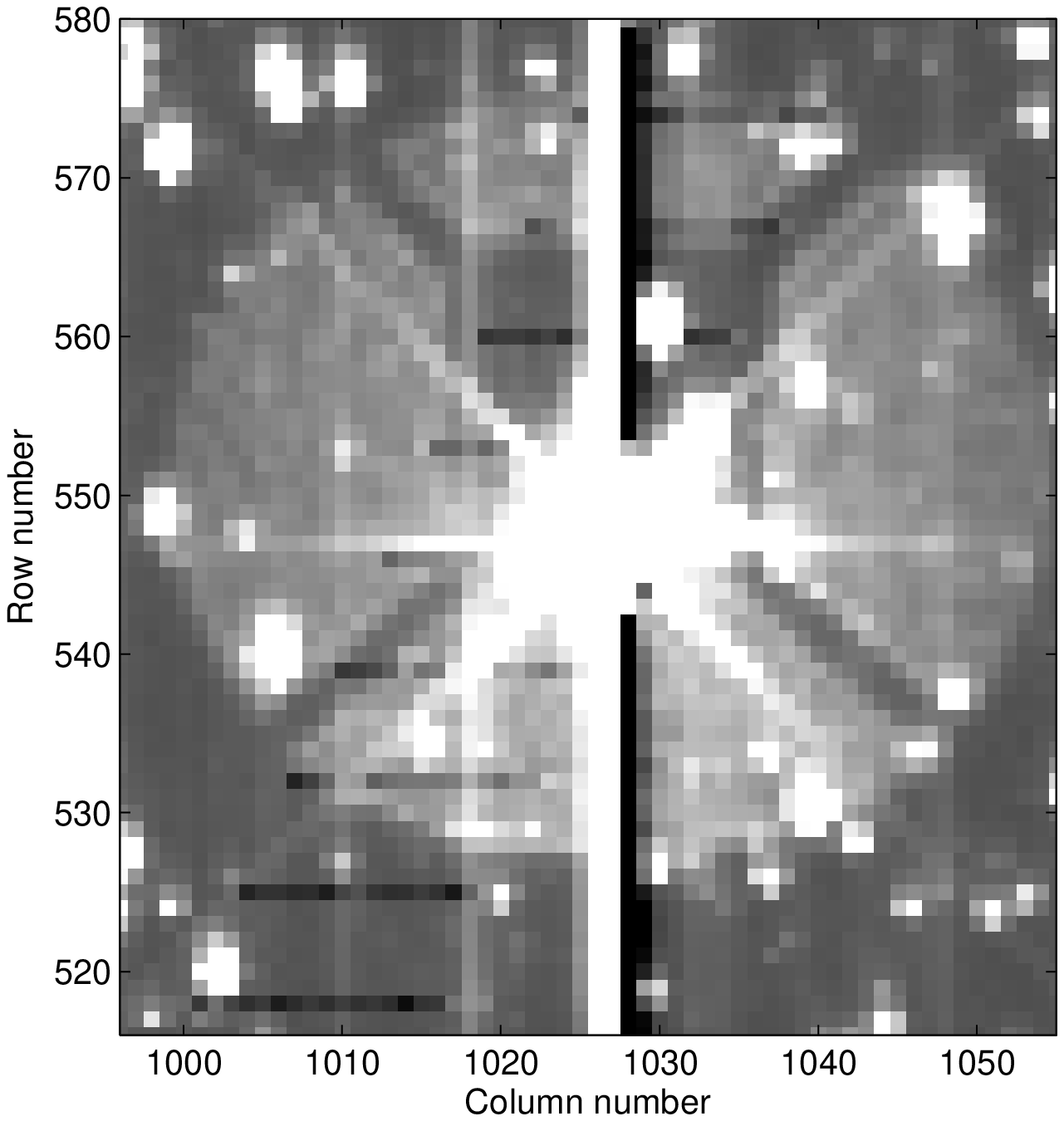}
\hspace{-0.6 in}
\includegraphics[scale=0.5,clip=true]{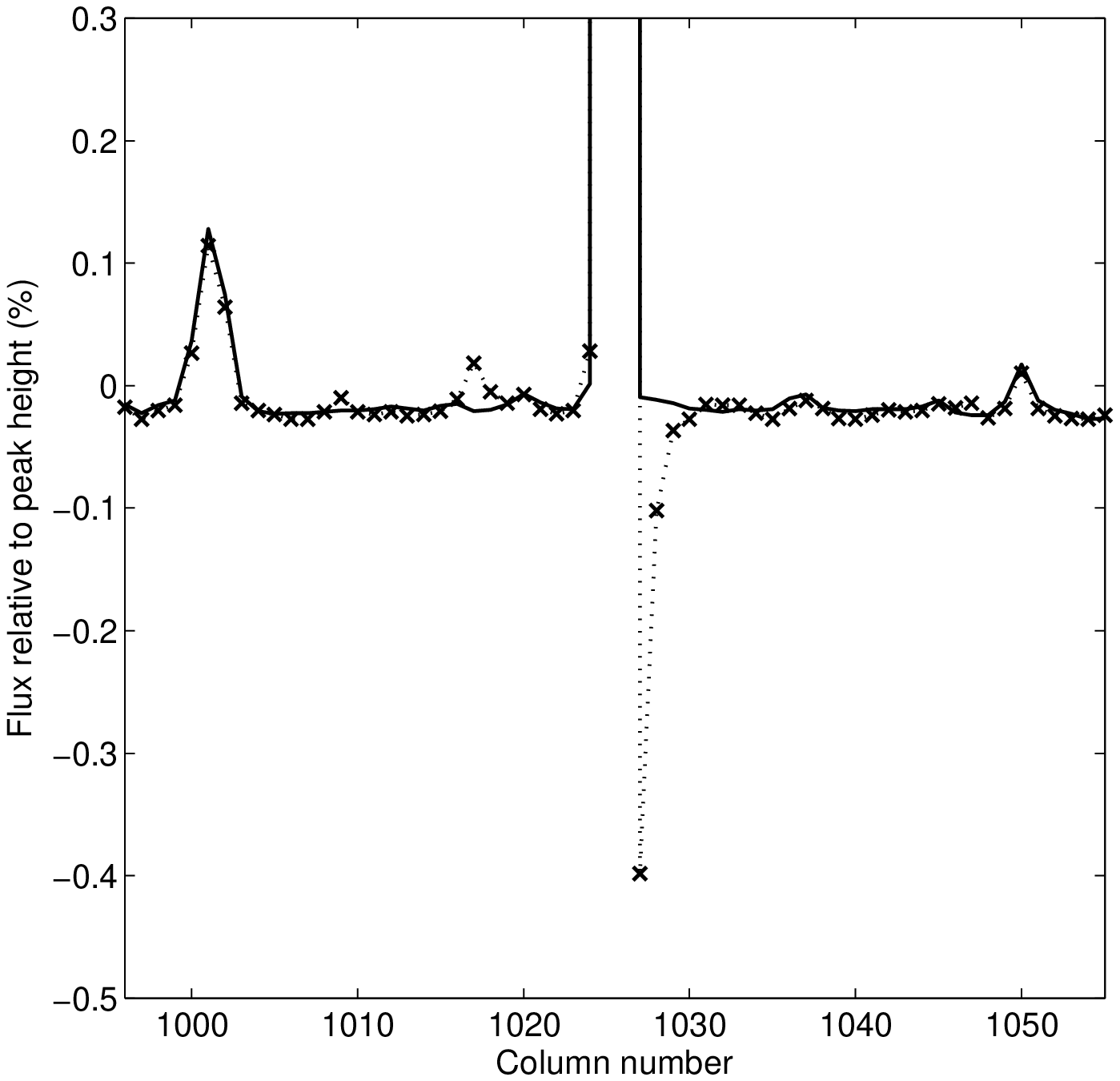}
\caption{Portion of a raw FFI of channel 11.1 near a saturating star 
that spills charge $\pm 30$ rows (left)
and the mean of three rows (521-523) cutting through
a saturated column (right), scaled to be relative to the saturation
peak height and expressed in percent. The dotted line with `x' symbols is from the 
raw FFI, and the solid line is from the calibrated FFI, demonstrating the efficacy 
of the undershoot correction \citep{jenkinsPipeline2010}. 
The undershoot level is $0.4\%$ for the first pixel and $0.1\%$ for the second.
This FFI (KPLR2009231194831) used 270 co-adds of 6.02 s integrations, 
for a total exposure time of 1625 s. 
The display uses a linear stretch from the $5^{\rm th}$ to $95^{\rm th}$ 
percentile. An optical ghost from the field flattener lens is clearly visible 
around this bright star. 
\label{fig:undershoot}}
\end{figure}

Testing of the LDE signal chain on star-like images revealed a large 
signal-dependent trailing undershoot in the video.  
The primary cause was 
traced to the use of a bipolar-input AD8021 operational amplifier 
in the correlated double-sample circuit.  
Corrective action to replace this amplifier with 
the junction field effect transistor-input AD8065 in all video 
channels resulted in greatly 
reducing this artifact, from an 
initial 2\% amplitude and 12 pixels duration to a median amplitude of 0.34\% 
and 3 pixels duration, as measured in flight data 
(Figure~\ref{fig:undershoot}). The
undershoot distortion can be modeled as an invertible, linear,
shift-invariant digital filter,
meaning we can correct the pixel values provided we have enough
pixels upstream (lower column numbers) of the pixel of interest. 
To this end, a column of pixels is prepended to each target 
aperture, and two targets were added
to each channel to monitor the undershoot response to both an impulse
and a step change. 
An undershoot correction is included in the pixel
calibration pipeline \citep{jenkinsPipeline2010}.

\subsection{FGS Clocking Cross Talk}
\label{sec:tvfx}

Cross talk from the FGS clocks to the 
science CCD video signals injects a complex pattern into the bias 
image of every science channel with an amplitude up to 
20~DN read$^{-1}$ \citep{argabrightSpie2008}. 
Because the FGS and science CCDs share the 
same master clock, the pattern is spatially fixed; however, the amplitude of the 
cross talk is dependent on the temperature of the LDE. The 
cross talk has three distinct components based on the state of the FGS CCDs as 
the science pixel is read out (see Figure~\ref{fig:ffi}): 
FGS CCD frame transfer, parallel transfer, and 
serial transfer (which shows no cross talk).
Approximately 20\% of targets have at least one of the parallel
or frame-transfer cross talk pixels in their aperture. Without mitigation,
the cross talk introduces a small time-varying bias into a target's flux time
series as the LDE temperature changes with orbital
position.

In order to measure the cross talk thermal dependence, 
a series of dark FFIs were taken at different temperatures during 
commissioning. Both the levels and their thermal dependence 
are consistent with ground characterizations. 
The signal in each cross talk pixel 
type can be modeled as an offset from the bias level that is 
linear in time and LDE board temperature \citep{vancleveKIH2009}.
This model removes the cross talk effect to the level of the read noise
for all but a few pixel types on the most affected channels. 
The temporal component
decreased significantly during commissioning with a damping time
on the order of a week, indicative of a transient
effect such as out-gassing. 
With the dust cover on, the focal plane was warmer during the 
dark data collection than 
during science operations, so the
model is used to extrapolate the measured cross talk levels to 
those we see during operations, allowing us to generate a 
bias image, or ``two dimensional (2D) black'' for calibration that includes the 
clocking cross talk.

\subsection{2D Black Artifacts}
Ground tests revealed several artifacts associated with the 2D 
structure of the dark images. 
Their extent on the focal plane was observed to be small, as was the likelihood 
that they would worsen or spread.  The 
dark images obtained in flight are completely consistent with ground 
test results. Nevertheless, 
the analysis pipeline does not yet include a means 
of identifying changes in these features, so time series that
are released with the current processing version may contain 
artifact features at signal levels discussed below. 

\subsubsection{High-frequency Oscillations}
A temperature-sensitive amplifier oscillation at  $>$1~GHz was 
detected in some CCD video channels during the artifact
investigation.  We suspect that the origin of this oscillation 
is from the AD8021 operational amplifiers used extensively in the 
video signal chain, 
which may show subtle layout-dependent instability when used at 
low gains. The oscillation's frequency range, rate of change, 
and pattern among the channels matched closely those 
characteristics in the dark images, 
strongly suggesting that the artifact is a \moire 
pattern generated by sampling the high-frequency oscillation at the $3 
\rm{MHz}$ serial pixel clocking rate. Since the characteristic source 
frequency drifts with time and temperature of the electronic components 
by as much as $500  \rm{kHz/ ^\circ C}$, the signal from a given pixel 
in a series of dark images has a time varying signature.  This signature 
may be highly correlated with neighboring pixels and yet poorly 
correlated with slightly more distant pixels.  When the oscillation 
frequency is a harmonic of the serial clocking frequency, a DC shift 
occurs producing a horizontal band offset from the mean bias-level
in the image.  As the frequency drifts with 
temperature, the point on the image where this DC shift occurs moves up 
or down from sample-to-sample, producing a rolling band.

Forty-six of the 84 readout channels have never exhibited this behavior and an 
additional 9 channels have thus far not exhibited this behavior at a 
detectable level in flight.  
Typically, in the remaining 29 channels, $ 
20\%$ of the FOV exhibits the \moire 
pattern with peak-to-peak amplitudes $>0.1$ DN read$^{-1}$ pixel$^{-1}$, 
while another $ 18\%$  exhibits between 
$0.02$ and 0.1 DN read$^{-1}$ pixel$^{-1}$.  
Two output channels, 9.2 and 17.2, 
typically exhibit $>0.1$ DN read$^{-1}$ pixel$^{-1}$ \moire pattern peak-to-peak 
amplitudes over their entire fields-of-view.  The resulting total FOV 
fraction typically affected by \moire signal at or above a level
of 0.1 DN read$^{-1}$ pixel$^{-1}$ 
(0.02 DN read$^{-1}$ pixel$^{-1}$) is 9\% (15\%).
For comparison, 
0.1 DN read$^{-1}$ pixel$^{-1}$ is the change in signal per pixel in a typical 
$12^{\rm th}$ magnitude star aperture for an Earth-size planet transit. 
We have adopted the conservative threshold of 
0.02 DN read$^{-1}$ pixel$^{-1}$, or roughly $1/4$ of the per pixel signal 
from an Earth-size transit, in order to put an upper bound on the
potential impact from image artifacts. While the \moire amplitude 
per pixel in these channels is significant, 
how the artifact affects our ability to detect small planets 
depends on its frequency, sum within a target aperture, and 
variations over time-scales of interest to transit detection. 
Based on the first 33.5 days of data from science operations, 
\citet{jenkinsLC2010} find the instrument is meeting the 6-hour 
precision requirement across the focal plane for the quietest 
30\% of stars. The two worst \moire channels, 9.2 and 17.2, 
exhibit a $\sim$20\% increase in 6-hour noise over the focal 
plane average at $12^{\rm th}$ magnitude, as measured by 
the standard deviation of 6-hour binned flux time series. 
Such an increase is small compared with the factor of 1.5 
spread in the distribution of dwarf star precision at 
$12^{\rm th}$ magnitude \citep{jenkinsLC2010}. 

\subsubsection{Scene-dependent Artifacts}
As a consequence of the sensitivity of the oscillating LDE component to 
temperature, the thermal transient introduced during readout by 
the signal from a bright star causes additional localized changes in 
bias-level (Figure~\ref{fig:ffi}).
These scene dependent artifacts persist over a range of hundreds of 
pixels in the columns  following bright stars.  In the 29 output 
channels exhibiting the \moire pattern, approximately $ 20\%$ of 
the FOV may be affected by these artifacts above the 0.02 
DN read$^{-1}$ pixel$^{-1}$ level. 

A second scene dependent effect is observed on all channels.  
A bright star produces a strong 
undershoot signal as discussed above, but the undershoot signal extends 
for hundreds of pixels at a very low level rather than the 20 pixels 
currently used for the inverse filter.  This extended undershoot signal 
varies in proportion to variations in the source star's light curve. 
We estimate that  $ 36\%$ of CCD rows contain stars bright 
enough to introduce extended undershoot signals which suggests that 
$20\%$ of pixels are at risk to be affected by this artifact.  Of these, 
we conservatively estimate that $ 50\%$ are actually affected above the 
0.02 DN read$^{-1}$ pixel$^{-1}$ level. This means $ 10\%$ of the total 
FOV is subject to this effect.

\subsubsection{Start-of-line Ringing}
A transient signal initiated at the onset of serial clocking of each row 
is well-modeled as a series of $\sim$5 superimposed, slightly under-damped 
oscillations which extend for roughly 150 pixels. 
This start-of-line ringing is evident on all output 
channels, but shows much less thermal sensitivity than 
the oscillations discussed previously.  The initial 
amplitude of the oscillations is $>$1~DN read$^{-1}$ pixel$^{-1}$; however, the 
pattern is static, so the current processing algorithms adequately 
remove the artifact to accuracies $<$0.02~DN read$^{-1}$ pixel$^{-1}$. 
We do not expect this artifact to impact photometric precision.

\section{Summary}
We have used commissioning and the first month of science operations to 
characterize \keplers instrument performance. The basic properties of the 
photometer are unchanged from ground testing. Image artifacts are consistent with 
ground observations, though star light creates scene dependence 
of the \moire pattern signal, complicating mitigation plans. Corrections in the current 
analysis pipeline for static FGS clocking cross talk, LDE undershoot, and start-of-line 
ringing are found to remove these artifacts to a level sufficient to meet \keplers 
precision requirements. We are currently testing mitigations for the 
thermally varying FGS cross talk, \moire pattern, and extended undershoot, which 
will subsequently be implemented in the analysis pipeline. Results indicate that we 
can reduce the levels of these artifacts to $<0.02$ DN read$^{-1}$ pixel$^{-1}$ over 
80\%--90\% of the focal plane, with the remainder flagged for use in subsequent 
analysis steps \citep{jenkinsPipeline2010}. The photometer is currently providing 
measurements of unprecedented precision and time coverage, giving us great 
confidence that the {\it Kepler Mission} will meet its science goals and make many 
unexpected discoveries.

\acknowledgments
We gratefully acknowledge the years of work by the many hundred members 
of the \kepler Team who conceived, designed, built, and now operate this
wonderful mission. Funding for this Discovery mission is provided by 
NASA's Science Mission Directorate.

{\it Facilities:} \facility{{\em Kepler}}.


\begin{thebibliography}{8}
\expandafter\ifx\csname natexlab\endcsname\relax\def\natexlab#1{#1}\fi

\bibitem[{{Argabright} {et~al.}(2008){Argabright}, {VanCleve}, {Bachtell},
  {Hegge}, {McArthur}, {Dumont}, {Rudeen}, {Pullen}, {Teusch}, {Tennant}, \&
  {Atcheson}}]{argabrightSpie2008}
{Argabright}, V.~S., {et~al.} 2008, Proc. SPIE, 7010, 70102L

\bibitem[{{Borucki} {et~al.}(2010){Borucki}, {Koch}, {Basri}, {Batalha},
  {Brown}, {Caldwell}, {Christensen-Dalsgaard}, {Cochran}, {Devore}, {Dunham},
  {Dupree}, {Gautier}, {Geary}, {Gilliland}, {Gould}, {Howell}, {Jenkins},
  {Kondo}, {Latham}, {Marcy}, {Meibom}, {Kjeldsen}, {Lissauer}, {Monet},
  {Morrison}, {Sasselov}, {Tarter}, Boss, Brownlee, Owen, Buzasi,
  {Charbonneau}, {Doyle}, Fortney, Ford, {Holman}, {Seager}, Steffen, Welsh,
  {Rowe}, Anderson, Buchhave, Ciardi, Walkowicz, Sherry, Horch, Isaacson,
  Everett, {Fischer}, {Torres}, {Johnson}, Endl, MacQueen, {Bryson}, {Dotson},
  {Haas}, {Kolodziejczak}, {Van Cleve}, {Chandrasekaran}, {Twicken}, V.,
  {Clarke}, {Allen}, {Li}, {Wu}, {Tenenbaum}, Anderson, Verner, Bruhweiler,
  Barnes, \& Prsa}]{boruckiScienceRev2010}
{Borucki}, W.~J., {et~al.} 2010, Science, submitted

\bibitem[{{Bryson} {et~al.}(2010){Bryson}, {Tenenbaum}, {Jenkins},
  {Chandrasekaran}, {Klaus}, {Caldwell}, {Gilliland}, {Haas}, {Dotson}, {Koch},
  \& {Borucki}}]{brysonPrf2010}
{Bryson}, S.~T., {et~al.} 2010, \apjl, {\em this issue}

\bibitem[{{Haas} {et~al.}(2010){Haas}, {Batalha}, {Bryson}, {Caldwell},
  {Dotson}, {Hall}, {Jenkins}, {Klaus}, {Koch}, {Kolodziejczak}, {Middour},
  {Smith}, {Sobeck}, {Stober}, {Thompson}, \& {Van Cleve}}]{haasSciops2010}
{Haas}, M.~R., {et~al.} 2010, \apjl, {\em this issue}

\bibitem[{{Jenkins} {et~al.}(2010{\natexlab{a}}){Jenkins}, {Caldwell},
  {Chandrasekaran}, {Twicken}, {Bryson}, V., {Clarke}, {Li}, {Allen},
  {Tenenbaum}, {Wu}, {Klaus}, {VanCleve}, {Dotson}, {Haas}, {Gilliland},
  {Koch}, \& {Borucki}}]{jenkinsLC2010}
{Jenkins}, J.~M., {et~al.} 2010{\natexlab{a}}, \apjl, {\em this issue}

\bibitem[{{Jenkins} {et~al.}(2010{\natexlab{b}}){Jenkins}, {Caldwell},
  {Chandrasekaran}, {Twicken}, {Bryson}, V., {Clarke}, {Li}, {Allen},
  {Tenenbaum}, {Wu}, {Klaus}, {Middour}, {Cote}, {McCauliff}, {Girouard},
  {Gunter}, {Wohler}, {Sommers}, {Hall}, {Wu}, {Van Cleve}, {Pletcher},
  {Dotson}, {Haas}, {Gilliland}, {Koch}, \& {Borucki}}]{jenkinsPipeline2010}
---. 2010{\natexlab{b}}, \apjl, {\em this issue}

\bibitem[{{Koch} {et~al.}(2010){Koch}, {Borucki}, {Basri}, {Batalha}, {Brown},
  {Caldwell}, {Christensen-Dalsgaard}, {Cochran}, {DeVore}, {Dunham},
  {Gautier}, {Geary}, {Gilliland}, {Gould}, {Jenkins}, {Kondo}, {Latham},
  {Lissauer}, {Marcy}, {Monet}, {Sasselov}, Boss, Brownlee, {Caldwell},
  {Dupree}, {Howell}, {Kjeldsen}, {Meibom}, {Morrison}, Owen, Reitsema,
  {Tarter}, {Bryson}, {Dotson}, {Gazis}, {Haas}, {Kolodziejczak}, {Rowe},
  {VanCleve}, {Allen}, {Chandrasekaran}, {Clarke}, {Li}, V., {Tenenbaum},
  {Twicken}, \& {Wu}}]{kochMissionDesign2010}
{Koch}, D.~G., {et~al.} 2010, \apjl, {\em this issue}

\bibitem[{{Van Cleve} \& {Caldwell}(2009)}]{vancleveKIH2009}
{Van Cleve}, J., \& {Caldwell}, D.~A. 2009, Kepler Instrument Handbook, KSCI
  19033-001 (Moffett Field, CA: NASA Ames Research Center), \tt{http://archive.stsci.edu/kepler/}

\end{thebibliography}

\clearpage

\clearpage

\begin{deluxetable}{lcrrr}
\tabletypesize{\scriptsize}
\tablecaption{Detector Properties Summary \label{tab:dpp}}
\tablewidth{0pt}
\tablehead{
\colhead{Parameter} & \colhead{Where Measured} & \colhead{Minimum} & 
\colhead{Maximum} & \colhead{Median}
}
\startdata
Read noise (e$^{-}$ read$^{-1}$) & Flight & 81  & 307 [149]\tablenotemark{a} & 
95 \\
Dark  current (e$^{-}$ pix$^{-1}$ s$^{-1}$) & Flight & 
-8.1[-3.2]\tablenotemark{a} & 7.5 & 0.25 \\
QE at 600 nm & Ground & 0.81 & 0.92 & 0.87 \\
Gain (e$^{-}$ DN$^{-1}$) & Ground & 94 & 120 & 112 \\
Saturation (Kepler Mag)\tablenotemark{b} & Flight & 11.6 & 10.3 & 11.3 \\
PRNU\tablenotemark{c} (\%) & Ground & 0.82 & 1.20 & 0.96 \\
LDE Undershoot (\%) & Flight & 0.08 & 1.92 & 0.34 \\
\enddata
\tablecomments{Minimum, maximum, and median are taken across all 84 
channels of the focal plane.}
\tablenotetext{a}{Flight measurements of read noise and dark current are 
contaminated by image artifacts for several of the noisiest channels 
(\S~\ref{sec:instartifacts}).  Values for non-artifact channels are given in 
square brackets, where different.}
\tablenotetext{b}{Saturation range is calculated using the measured QE
variations, a vignetting model, and the observed central pixel flux
fraction range of 0.28--0.64.}
\tablenotetext{c}{PRNU is a measure of local pixel response variation and 
does not include large-scale optical effects such as vignetting 
(\S~\ref{sec:stdccd}).}
\end{deluxetable}


\begin{deluxetable}{lrr}
\tablecaption{Field of View at Risk from Artifacts.\label{tab:artifact}}
\tablewidth{0pt}
\tablehead{
 & \colhead{Potential Extent} & \colhead{FOV at Risk} \\
 \colhead{Artifact} & \colhead{on FOV\tablenotemark{a}} (\%) & 
 \colhead{$>0.02$ DN/read\tablenotemark{a}} (\%)
 }
\startdata
LDE undershoot & 100 & 0 \\
FGS cross talk & 20  & 0 \\
Moir\'e pattern & 45 & 15\tablenotemark{b} \\
Scene dependent moir\'e & 45 & 7\tablenotemark{b} \\
Extended undershoot & 100 & 10\tablenotemark{b} \\
\enddata
\tablenotetext{a}{``Potential extent'' indicates the fraction of the FOV where the artifact could potentially be seen. ``FOV at risk'' indicates the fraction where we could see time varying signals at or above 0.02 DN read$^{-1}$ after existing and planned mitigations in the analysis pipeline. See the text for discussion.}
\tablenotetext{b}{The FOV values do not add since artifacts overlap. Scene dependent moir\'e pattern drift adds $\sim4\%$ unique FOV and extended undershoot adds $\sim8\%$.}
\end{deluxetable}

\end{document}